\documentstyle[12pt]{article}
\title{A Note on Dual Superconductivity and Confinement.}
\author{Wifredo Garc\'{\i}a Fuertes \\Departamento de F\'\i sica, Facultad de Ciencias, \\ Universidad de Oviedo \\ SPAIN\\and\\Juan Mateos Guilarte \\Departamento de F\'\i sica, Facultad de Ciencias, \\ Universidad de Salamanca \\ SPAIN}
\date{}
\newcommand{\beq}{\begin{equation}}
\newcommand{\eeq}{\end{equation}}
\newcommand{\bdm}{\begin{displaymath}}
\newcommand{\edm}{\end{displaymath}}
\begin{document}
\maketitle
\begin{abstract}
Electric self-dual vortices arising as BPS states in the strong coupling limit of
N=2 supersymmetric Yang-Mills theory, softly broken to $N=1$, are reported.
\end{abstract}
\section{.} 
About twenty years ago, 't Hooft and Mandelstam \cite{bib:1}, \cite{bib:2} proposed a qualitative description of the quark confinement phenomenon based on an analogy with superconductivity.  According to this interpretation, the QCD vacuum behaves as a ``dual" superconductor, a material where magnetic charge condensation and the electric Meissner effect take place. At that time, it was not known how local quantum field operators that create or annihilate magnetically charged quanta could arise in a non-abelian gauge theory and the 't Hooft-Mandelstam picture remained as a ``natural" but descriptive scheme of quark confinement. 

Pursuing another brilliant idea, the Olive-Montonen electro-magnetic duality conjecture \cite{bib:3}, Seiberg-Witten \cite{bib:4} showed that in the family of quantum field theories describing $N=2$ supersymmetric Yang-Mills systems at a specific quantum vacuum, the 't Hooft insight does indeed hold. In certain patch of the quantum moduli space of vacua the local fields carry and/or are coupled to magnetic charge. The conventional local fields of $N=2$ SUSY YM theory appropriately describe the system in another patch and  are related with those previously mentioned
by a non-local transformation suggested by electro-magnetic duality. The Olive-Montonen conjecture is subsequently proven in a far from trivial manner as 
a duality of the Wilsonian effective theories at weak and strong coupling, and there is a third patch to completely cover the quantum moduli space of vacua where the local fields bring both electric and magnetic charge.  

Moreover, by softly breaking $N=2$ to $N=1$  supersymmetry, Seiberg and Witten were also able to show that magnetic charge condensation and the electric Higgs mechanism are both displayed at the strong coupling limit. It therefore seems that the 't Hooft-Mandelstam scenario is analytically verified in softly broken $N=2$ QCD, but there is an important element that has not yet been discussed, despite the enormous amount of research devoted to this subject in recent years.
How do electric flux lines arise at the strong coupling limit in the theory? From the superconductivity analogy, we recognize electric vortices as the crucial entities for electric charge confinement. We shall show that self-dual electric vortices exist and that electric charge confinement  therefore occurs in the system at the critical point between the Type I and Type II phases of a dual superconductor.

In $N=2$ SUSY YM there is a renormalization-invariant scale $\Lambda^2=\mu^2{\rm exp}\{-\frac{4\pi^2}{g^2(\mu)}\}$ dynamically generated by quantum effects such that $g(\Lambda^2)=\infty$; i.e. the limit of the renormalization point going to $\Lambda^2$ is the strong coupling limit of the system. If $u$ is a holomorphic coordinate in the quantum moduli space of vacua it has been proved that when $u=\Lambda^2$ the Wilsonian effective action is $N=2$ supersymmetric and tantamount to dual QED with massless electrons, see Reference \cite{bib:5} for excellent reviews. Therefore, the abelian effective action includes two contributions:
\begin{enumerate}
\item 
\beq
S_W=\int d^4x\{\frac{1}{4}\int[d^2\theta W_\alpha^DW^\alpha_D+{\rm h.c.}]+\int d^2\theta d^2\bar{\theta}\chi_D^\dagger\chi_D\}\label{eq:1}
\eeq
due to the $N=1$ chiral $\chi_D$ and vector $V_D$ multiplets. The fields entering (\ref{eq:1}) are the dual photon and Higgs field and their supersymmetric partners. Performing the Berezin Integration we have
\begin{eqnarray}
S_W&=&\int d^4x\{-\frac{1}{4}F_{\mu\nu}^DF^{\mu\nu}_D-i\bar{\lambda_D}\bar{\sigma}^\mu\partial_\mu\lambda_D+\frac{D^2}{2}+\nonumber\\ &+&\partial_\mu\phi^*_D\partial^\mu\phi_D-i\bar{\psi}_D\bar{\sigma}^\mu\partial_\mu\psi_D+F^*_DF_D\}
\end{eqnarray}
where $\lambda_D$, $\psi_D$ are Weyl spinor fields and $D$ and $F_D$ are auxiliary fields. $\sigma^\mu=({\bf 1}_2,\sigma^i)$  and $ \bar{\sigma}^\mu=({\bf1}_2,-\sigma^i) $ are vectors of $2\times 2$ matrices where $\sigma^i$ are the Pauli matrices. The subscript $D$ means dual and by this we mean that $A_D^\mu$ is the vector potential associated with the group $U(1)_D$, the dual of $U(1)$, the maximal torus of the $SU(2)$ gauge group; the remaining $\phi_D$, $\lambda_D$ and $\psi_D$ fields form the $N=2$ multiplet with $A_\mu^D$. Although $\phi_D$ is a complex scalar field, it is neutral with respect to the $U(1)_D$ charge. $\lambda_D$ and $\psi_D$ do not carry magnetic charge either; thus, there is no minimal coupling of $\phi_D$, $\lambda_D$ and $\psi_D$ to $A_\mu^D$.
\item Because at $u=\Lambda^2$ the BPS magnetic monopoles become massless, a term of the form
\beq
S_M=\int d^4x \int d^2\theta d^2\bar{\theta}(\tilde{M}^\dagger e^{2\bar{g}_DV_D}\tilde{M}+M^\dagger e^{-2\bar{g}_DV_D}M).
\eeq
should be added to the Wilson effective action. $M$ and $\tilde{M}^\dagger$ form the $N=2$ matter multiplet of ``magnetically'' charged fields: the quanta from the complex scalar fields $\phi_m$ and $\tilde{\phi}_m$ are, respectively, the scalar dual electrons and scalar dual positrons while the Weyl spinor fields $\psi_m$ and $\tilde{\psi}_m$ have dual electrons and positrons as quanta. It is important to notice that $S_M$ is a ``classical'', Ginzburg-Landau type, action; all the quantum fluctuations of $k^2>0$ for the magnetically charged particles, as well as the $\chi_D$ quanta, have been integrated out in $S_W$. We are left with classical monopole, chiral and vector fields: $M$, $\tilde{M}$, $\chi_D$ and $V_D$ which should be interpreted as the expectation values of the partner quantum fields at the possible quantum states. Thus, $\bar{g}_D$ is the coupling constant
of the $M$ and $\tilde{M}^\dagger$ classical fields to $V_D$; through wave particle duality $\bar{g}_D$ is connected to the particle magnetic charge: $\hbar\bar{g}_D=g_D(0)$. The strong coupling limit becomes the classical limit, $g_D(0)=0$ if $\bar{g}_D\neq 0$, for the magnetically charged quanta.
\end{enumerate}
N=2 supersymmetry, unbroken by the quantization procedure, also requires the addition of a Yukawa term:
\beq
S_Y=\sqrt{2}\bar{g}_D\int d^4x[\int d^2\theta\chi_D M\tilde{M}+{\rm h.c.}].\label{eq: am}
\eeq
The coupling constant is fixed by supersymmetry and Berezin integration allows for the explicit expressions,
\begin{eqnarray}
S_M&=&\int d^4x\{(D_\mu\phi_m)^*D^\mu\phi_m+D_\mu\tilde{\phi}_m(D^\mu\tilde{\phi}_m)^*+F^*_mF_m+\tilde{F}_m^*\tilde{F}_m+\nonumber\\ &+&\bar{g}_DD(\phi^*_m\phi_m-\tilde{\phi}_m^*\tilde{\phi}_m)-i(\bar{\psi}_m\sigma^\mu D_\mu\psi_m+\bar{\tilde{\psi}}_m\sigma^\mu D_\mu\tilde{\psi}_m)\}
\end{eqnarray}
where the covariant derivatives are
\begin{eqnarray}
D_\mu\phi_m=\partial_\mu\phi_m+i\bar{g}_DA_\mu^D\phi_m,&\;\;\;\;\;\;&D_\mu\psi_m=\partial_\mu\psi_m+i\bar{g}_DA_\mu^D\psi_m\\D_\mu\tilde{\phi}_m=\partial_\mu\tilde{\phi}_m-i\bar{g}_DA_\mu^D\tilde{\phi}_m&\;\;\;\;\;\;&D_\mu\tilde{\psi}_m=\partial_\mu\tilde{\psi}_m-i\bar{g}_DA_\mu^D\tilde{\psi}_m
\end{eqnarray}
and
\begin{eqnarray}
S_Y&=&\sqrt{2}g_D\int d^4x[\{F_D\phi_m\tilde{\phi}_m+2\psi_D(\tilde{\psi}_m\phi_m+\psi_m\tilde{\phi}_m)+\phi_D(\phi_m\tilde{F}_m+\tilde{\phi}_mF_m)\nonumber\\ &+&{\rm h.c.}\}-\{\phi_m^*\psi_m\lambda_D-\phi_m\bar{\psi}_m\bar{\lambda}_D+\tilde{\phi}_m\bar{\tilde{\psi}}_m\bar{\lambda}_D-\tilde{\phi}^*\psi_m\lambda_D\}].
\end{eqnarray}

Therefore, 
\beq
S_{eff}=S_W+S_M+S_Y
\eeq
is the ``classical'' effective theory coming from $N=2$ SUSY YM at the strong coupling limit.
\section{.}
Adding a mass term in the original non-abelian theory
\beq
{\cal L}[\mu]=\mu\int d^2\theta[{\rm Tr}\Phi^2+{\rm h.c.}]=-{\rm Tr}(\mu\phi^*\phi+i\mu(\psi\psi+\bar{\psi}\bar{\psi}))
\eeq
$N=2$ supersymmetry is softly broken to $N=1$. A dramatic change occurs in the effective theory: ${\rm Tr}\Phi^2$ becomes an abelian superfield $U(y,\theta)$ which in the monopole patch is a functional of $\chi_D$,
\beq
U(\chi_D)=U(\phi_D)+\sqrt{2}\theta\psi_DU^\prime(\phi_D)+\theta^2(U^\prime(\phi_D)F_D-\frac{1}{2}U^{\prime\prime}(\phi_D)\psi_D\psi_D).
\eeq
There is another contribution at $u=\Lambda^2$ to the ``classical" effective action,
\beq
S_\mu=\mu\int d^4x \{U^\prime(\phi_D)F_D-\frac{1}{2}U^{\prime\prime}(\phi_D)\psi_D\psi_D+{\rm h.c.}\}.
\eeq
New features due to $S_\mu$ come essentially from the constant modes. The effective action reduces to the effective potential
\beq
V_{eff}=\sqrt{2}\bar{g}_D\chi_DM\tilde{M}+\mu U(\chi_D)+{\rm h.c.}.
\eeq
The expectation values of $M$, $\tilde{M}$ and $\chi_D$ at the Lorentz invariant quantum ground states
\beq
\langle M(x)\rangle=\langle\phi_m(x)\rangle=m,\ \langle \chi_D(x)\rangle=\langle\phi_D(x)\rangle=a_D,\ \langle \tilde{M}(x)\rangle=\langle\tilde{\phi}_m(x)\rangle=\tilde{m}
\eeq
are given by the minima of $V_{eff}$
\beq
\sqrt{2}\bar{g}_Dm\tilde{m}+\mu U^\prime (a_D)=0,\;\;\;\;\;\; a_D\tilde{m}=a_Dm=0\label{eq:34}
\eeq
There is also the constraint $|m|=|\tilde{m}|$ coming from the $D$-terms. When $\mu=0$, $m=\tilde{m}=0$ and $a_D\in{\bf C}$ solve (\ref{eq:34}); the moduli space of vacua of the $N=2$ theory arises. If $\mu U^\prime(0)\neq 0$ the vacuum manifold is different:
\beq
a_D=0,\;\;\;\;\; m\tilde{m}=|m|^2e^{i(\alpha+\tilde{\alpha})}=|\frac{\mu}{\bar{g}_D\sqrt{2}}U^\prime(0)|e^{i\beta}
\eeq
is the solution of (\ref{eq:34}). Only the sum of the phases of $m$ and $\tilde{m}$ is fixed, $\beta =\alpha+\tilde{\alpha}$ and there is a ``circle" of vacua. (Massless) magnetic monopoles, dual s-electrons, condense: spontaneous symmetry breakdown of the dual abelian group arises
from identifying the vacuum manifold as the orbit of the $U(1)_D$ action.

Expanding around the expectation values of the $M$ and $\tilde{M}$ fields
\begin{eqnarray}
M(y,\theta)&=&m+h_m(y)+\sqrt{2}\theta\psi_m(y)+\theta^2F_m(y)\\
\tilde{M}(y,\theta)&=&\tilde{m}+\tilde{h}_m(y)+\sqrt{2}\theta\tilde{\psi}_m(y)+\theta^2\tilde{F}_m(y).
\end{eqnarray}
scalar and vector mass terms arise in the lagrangian:
\beq
{\cal L}_m=\bar{g}_D^2|m|^2(|h_m|^2+|\tilde{h}_m|^2+A_\mu^DA_D^\mu)
\eeq
The ``electric" Higgs mechanism for the fermions, albeit in a more involved manner, also happens, see Reference \cite{bib:6}. The conditions for the existence of ``vortices" are met, both from topological and dynamical viewpoints. 

The ``bosonic'' part of the effective action is,
\begin{eqnarray}
S_{eff}^B&=&\int d^4x\{-\frac{1}{4}F_{\mu\nu}^DF^{\mu\nu}_D+\partial_\mu\phi_D^*\partial^\mu\phi_D+(D_\mu\phi_m)^*D^\mu\phi_m+(D_\mu\tilde{\phi}_m)^*D^\mu\tilde{\phi}_m+\nonumber\\ &+&\bar{g}_D D(|\phi_m|^2-|\tilde{\phi}_m|^2)+F_D^*F_D+[F_D((\sqrt{2}\bar{g}_D\phi_m\tilde{\phi}_m+\mu U^\prime(\phi_D))+{\rm h.c}]+\nonumber\\ &+&F_m^*F_m+\tilde{F}_m^*\tilde{F}_m+[\sqrt{2}\bar{g}_D\phi_D(\phi_m\tilde{F}_m+\tilde{\phi}_mF_m)+{\rm h.c}]\}
\end{eqnarray}
On shell, integrating out the auxiliary fields, this becomes $
S_{eff}^B=
\int d^4x\{T_{eff}^B-V_{eff}^B\}$

\begin{eqnarray}
V_{eff}^B&=&\frac{\bar{g}_D^2}{2}(|\phi_m|^2-|\tilde{\phi}_m|^2)^2+|\sqrt{2}\bar{g}_D\phi_m\tilde{\phi}_m+\mu U^\prime (\phi_D)|^2+\nonumber\\ &+&2\bar{g}_D^2|\phi_D|^2(|\phi_m|^2+|\tilde{\phi}_m|^2)\label{eq:21}
\end{eqnarray}
 Let us define a two-component field $S=\left(\begin{array}{cc}\phi_m\\\tilde{\phi}_m^*\end{array}\right)$; $\phi_m$ and $\tilde{\phi}_m^*$ transform under the $SU(2)_I$ group which act on the doublet of supercharges of the $N=2$ theory as the fundamental representation. We obtain:
\begin{eqnarray}
T_{eff}^B&=&-\frac{1}{4}F_{\mu\nu}^DF^{\mu\nu}_D+\partial_\mu\phi_D^*\partial^\mu\phi_D+(D_\mu S)^\dagger D^\mu S\label{eq:22}\\V_{eff}^B(\mu)&=&\frac{\bar{g}_D^2}{2}(4|\phi_D|^2+S^\dagger S)S^\dagger S+\nonumber\\+ \sqrt{2}\bar{g}_D&[&\mu^*U^*(\phi_D^*)S^\dagger\left(\begin{array}{cc}0&0\\1&0\end{array}\right)S+ {\rm h.c.}]+|\mu U^\prime (\phi_D)|^2\label{eq:23}
\end{eqnarray}
If there are minima of $S_{eff}^B$ which are space-time dependent, this means that there are quantum ground states such that the expectation values of $M$, $\tilde{M}$ and $\chi_D$ correspond to classical solutions of the field equations derived from (\ref{eq:22})-(\ref{eq:23}). The only consistent possibility at $u=\Lambda^2$ is $\langle\chi_D(x)\rangle =\langle\phi_D(x)\rangle=a_D=0$; from equation (\ref{eq:34}) and the second term in (\ref{eq:21}) we see that minima of $S_{eff}^B$ such that $\phi_m\tilde{\phi}_m(x)\neq m\tilde{m}$ are possible if and only if $U^{\prime\prime}(0)=0$. But this is indeed the case because when $u\rightarrow\Lambda^2$, $U(\chi_D)\simeq a_0\chi_D+\Lambda^2$ where $a_0$ is a constant, as one can check from the Seiberg-Witten solution
for the prepotential near $u=\Lambda^2$  and $u=\infty$, see \cite{bib:5}.

\section{.}
The effective action being abelian, we expect to find planar solitons; we look for minima of the effective energy density
\begin{eqnarray}
{\cal E}_{eff}^B(a_D=0)&=&\int d^2x\{\frac{1}{4}F_{ab}^DF_{ab}^D+(D_a\phi_m)^*D_a\phi_m+(D_a\tilde{\phi}_m)^*D_a\tilde{\phi}_m+\nonumber\\ &+&|\sqrt{2}\bar{g}_D\phi_m\tilde{\phi}_m+C|^2+\frac{\bar{g}_D^2}{2}(|\phi_m|^2-\tilde{\phi}_m|^2)^2\}
\end{eqnarray}
Here $a,b=1,2$ and we have defined $\mu U^\prime(0)=C=C_1+iC_2$. Following the method of Reference \cite{bib:7}, we restrict the search to the surface in the $N=2$   I-space  ${\bf C}^2$ defined by
\beq
\tilde{\phi}_m=-e^{i\beta}\phi_m^*   
\eeq ,
${\displaystyle \tan\beta=\frac{C_2}{C_1}}$ , together with the asymptotic behaviour
\beq
\lim_{r\rightarrow\infty}\phi_m(x_1,x_2)=m,\ \ \ \ \  \lim_{r\rightarrow\infty}\tilde{\phi}_m(x_1,x_2)=\tilde{m},\ \ \ \ \ |m|^2=|\tilde{m}|^2=\frac{|C|}{\sqrt{2}\bar{g}_D}
\eeq
Re-scaling the $\phi_m$-field to ${\displaystyle \phi_m=\frac{1}{2}\phi}$ the bosonic effective density, at this particular disc centred at the origin in ${\bf C}^2$, becomes
\beq
{\cal E}_{eff}^B(a_D=0,\beta)=\int d^2x\{\frac{1}{2}F_{12}^DF_{12}^D+\frac{1}{2}(D_a\phi)^*D_a\phi+\frac{\bar{g}_D^2}{8}(|\phi|^2-f^2)^2\}
\eeq
where ${\displaystyle f^2=2\sqrt{2}\frac{|C|}{\bar{g}_D}}$. The familiar energy density for the abelian Higgs model arises \cite{bib:8} and from the Bogomolnyi splitting
\begin{eqnarray}
{\cal E}_{eff}^B(a_D=0,\beta)&=&\frac{1}{2}\int d^2x\{[F_{12}^D\pm\frac{\bar{g}_D}{2}(f^2-|\phi|^2)]^2+|(D_1\pm iD_2)\phi|^2\}\mp\nonumber\\&\mp&\int d^2x\{\frac{\bar{g}_D}{2}(f^2-|\phi|^2)F_{12}^D+\frac{i}{2}(D_1\phi^* D_2\phi-D_2\phi^*D_1\phi)\},
\end{eqnarray}
it is seen that the absolute minima satisfy the first order equations:
\begin{eqnarray}
(D_1\pm iD_2)\phi=0\label{eq:30}\\
F_{12}^D=\mp\frac{\bar{g}_D}{2}(f^2-|\phi|^2)\label{eq:31}
\end{eqnarray}
because, up to a total divergence,
\begin{eqnarray}
{\cal E}_{eff}^B(a_D=0,\beta)&=&\int d^2x\{[F_{12}^D\pm\frac{\bar{g}_D}{2}(f^2-|\phi|^2)]^2+|(D_1\mp iD_2)\phi|^2\}\mp\nonumber\\&\mp&\int d^2x\{\frac{\bar{g}_D}{2}f^2F_{12}^D\}
\end{eqnarray}

 There are self-dual vortex solutions to (\ref{eq:30})-(\ref{eq:31})  having quantized, in this case electric, flux: ${\displaystyle \Phi_e=\int d^2xF_{12}^D=-\frac{2\pi}{\bar{g}_D}n}$, $n\in{\bf Z}$. Quantization of the flux is due to topological considerations and the
 energy density is finite and  easy to compute, see \cite{bib:8}:
\beq
{\cal E}_{eff}({\rm self-dual},\beta)=\pi f^2n ;
\eeq
the self-dual solutions are thus, electric flux tubes which appear at the critical point between Type I and Type II dual superconductivity. In the $N=2$ effective theory at strong coupling, if smoothly broken to $N=1$  as described above, there is a moduli space of self-dual electric vortices of dimension $2n$, parametrized by the vortex centres.

It is remarkable that there is a one-to-one correspondence between the set of values ${\displaystyle U^\prime(0)=\frac{C}{\mu}}$ and the set of trial surfaces to be chosen. As examples, we mention four special cases:
\begin{enumerate}
\item $C_2=0$, $C_1>0$. $\mu U^\prime(0)$, real positive. $\beta=0$ and $\displaystyle \tilde{\phi}_m=-\phi_m^*$.
\item $C_2=0$, $C_1<0$. $\mu U^\prime(0)$, real negative. $\beta=\pi$ and $\displaystyle \tilde{\phi}_m=\phi_m^*$.
\item $C_2>0$, $C_1=0$. $\mu U^\prime(0)$, purely imaginary. ${\displaystyle \beta=\frac{\pi}{2}}$ and $\tilde{\phi}_m=-i\phi_m^*$.
\item $C_2<0$, $C_1=0$. $\mu U^\prime(0)$, purely imaginary. ${\displaystyle \beta=-\frac{\pi}{2}}$ and $\tilde{\phi}_m=i\phi_m^*$.
\end{enumerate}
We have thus found that whatever the value of $\mu U^\prime(0)$ there exist stable electric vortices of finite energy density in the Ginzburg-Landau type of effective theory at the stake if a term softly  breaking the supersymmetry to $N=1$ is added. Alternatively, the vortex solutions can be seen as the expectation values of quantum fields at the quantum ground states at each topological sector, labeled by the number of ``quanta" of electric flux ${\displaystyle \frac{2\pi}{\bar{g}_D}}$. Notice that, if they are seen from a (3+1)-dimensional perspective, our self-dual vortices have a stringy character; they are line-like defects. In these circumstances the Wilson criterion of electric charge confinement is satisfied and  we thus confirm in a more direct way the result of Reference \cite{bib:9}, where the correct commutation relations for confinement of the non-local order-disorder operators were derived.

It is intriguing that at the strong coupling limit where the magnetically charged solitonic BPS states become fundamental quanta, other kind of electrically charged planar solitonic BPS states are generated. The BPS spectrum is given by the massive dual scalar electrons (magnetic monopoles) and their SUSY partners on one side. On the other side there are heavier electric vortices. To understand this, recall that BPS states are related to extended supersymmetry \cite{bib:10}. Apparently, the breaking of $N=2$ supersymmetry by the $U$ superfield should forbid the critical point from being reached. But there is still $N=1$  supersymmetry from the (3+1)-dimensional point of view that becomes $N=2$ supersymmetry when dimensional reduction to (2+1)-dimensions occurs. The price to be paid for having BPS states after explicitly breaking the $N=2$ supersymmetry is a reduction of their dimensionality. This is also related to the fact that the effective theory is abelian and the mass of the heavy objects, the electric vortices and their SUSY partners, comes from the central charge of the $N=2\ $ (2+1)-dimensional extended supersymmetry: the electric flux.

A stronger supersymmetry breaking from $N=2$ to $N=1$  as meant by choosing different coefficients of the kinetic terms for the vector and chiral multiplets, under the (unwarranted) hypothesis that an analogous bosonic effective action at strong coupling would appear, leads to either Type I or Type II phases of dual superconductivity. In that case, the $N=1$ (3+1)-dimensional supersymmetry is not reorganized as $N=2$ (2+1)-dimensional supersymmetry of the planar solitonic system.

\section{.}
Finally, we briefly discuss the case $\mu=0$. The $N=2$ effective theory in 3+1 dimensions leads to the energy density:
\beq
{\cal E}_{eff}^B(a_D=0)=\int d^2x\{\frac{1}{4}F_{ab}^DF^D_{ab}+(D_{a}S)^{\dagger}D_{a}S+\frac{\bar{g}_D^2}{2}(S^{\dagger}S)^2
\}\label{eq:43}
\eeq
The Bogomolny bound is reached by the solutions of
\begin{eqnarray}
F_{12}^D&=&\pm\bar{g}_D(S^{\dagger}S)\label{eq:44}\\
(D_1&\pm&iD_2)S=0\label{eq:45}
\end{eqnarray}
The self-duality equations have the wrong sign: there are no regular solutions to (\ref{eq:44})-(\ref{eq:45}) other than $S=F_{12}^D=0$. The system is ``frustrated": there are no electrically charged solitonic BPS states at the strong coupling limit. 

Such frustration may be relieved by passing to a curved background, see \cite{bib:11}. We choose  the Poincar\`e metric in the disc of radius c of the $(x_1, x_2)$-plane. A term such as
\beq
{\cal L}_R=-\frac{1}{2}R(S^{\dagger}S)\label{eq:49}
\eeq
arises in the $N=2$ supersymmetric effective Lagrangian with scalar curvature $ {\displaystyle R=-\frac{2}{c^2}}$ in this case. Passing to non-dimensional variables ${\displaystyle x_i\rightarrow\frac{x_i}{\bar{g}_Df}}$, $S\rightarrow fS$ and $A_a\rightarrow fA_a$ where ${\displaystyle f^2=\frac{1}{\bar{g}_D^2c^2}}$, the first order equations become:
\begin{eqnarray}
F_{12}^D&=&\mp\frac{4}{(1-(x_1^2+x_2^2))^2}(1-S^{\dagger}S)\label{eq:50}\\(D_1&\pm&iD_2)S=0,\label{eq:51}
\end{eqnarray}
if $(x_1^2+x_2^2)\leq1$.
The regular solutions, if $\tilde{\phi}_m=0$, are: 
\begin{eqnarray}
\phi_m^{(n)}(z)&=&\frac{2(1-|z|^2)}{1-f_n^*f_n(z)}\frac{f_n^{\prime}(z)}{|f_n^{\prime}(z)|}\label{eq:52}\\f_n(z)&=&\prod_{i=1}^{n}\frac{(z-a_i)}{(1-a_{i}^{*}z)},\label{eq:53}
\end{eqnarray}
where $|a_i|\leq1, \forall{i}, i=1,2,...,n $, correspond to self-dual vortices of quantized electric flux:
${\displaystyle \Phi_e=\frac{2\pi}{\bar{g}_D}(n-1)}$, see \cite{bib:12}. 
The phase $\Theta_m$ of $\phi_m$ is fixed as:
\beq
\Theta_m^{(n)}(z,\bar{z})= 2\sum_{i=1}^{n-1}\arg(z-z_0^{(i)})\label{eq:54}
\eeq
in terms of the zeros $z_0^{(i)}$ of $f_n^{\prime}(z)$. This choice renders the vorticial
vector field
\beq
A_z^{(n)}=-i\partial_{\bar{z}}{\rm log}\phi_m^{(n)}(z,\bar{z})\label{eq:55}
\eeq
single-valued.

There is therefore a moduli space
of electrically-charged heavy BPS states of dimension 2$n$, the dual counterpart of the BPS
magnetic monopoles which are point solitons at weak coupling. As suggested in Reference \cite{bib:13} the BPS spectra at strong and weak coupling have dual features: light quanta
are magnetically-charged while heavy objects are electric strings at the first limit. At weak coupling, light quanta and solitons are charged in the opposite way. A background of negative curvature is necessary at strong coupling to obtain this picture.


\begin{thebibliography}{aa}
\bibitem{bib:1}G. 't Hooft, Nucl. Phys. B153 (1979) 141.
\bibitem{bib:2}S. Mandelstam, Phys. Rep. 23C (1976) 245.
\bibitem{bib:3}C. Montonen and D. Olive, Phys. Lett. 72B (1977) 117.
\bibitem{bib:4}N. Seiberg and E. Witten, Nucl. Phys. B426 (1994) 19.
\bibitem{bib:5}A. Bilal, hep-th/9601007;  L. Alvarez-Gaum\'e and F. Hassan, Forst. Phys. 45 (1997) 159; W. Lerche, hep-th/961190.
\bibitem{bib:6}M. Di Pierro and K. Konishi, Phys. Lett. B388 (1996) 90.
\bibitem{bib:7}J. Mateos Guilarte, Nucl. Phys. B420 (1994) 315; W. Garc\'{\i}a Fuertes and J. Mateos Guilarte, Phys. Rev. D49 (1994) 6687.
\bibitem{bib:8}A. Jaffe and C. Taubes, ``Vortices and Monopoles", Birkhauser (1980).
\bibitem{bib:9}Chan Hong-Mo and Tsun Sheung Tsun, Phys. Rev. D53 (1996) 7293.
\bibitem{bib:10}D. Olive and E. Witten, Phys. Lett. B78 (1978) 97.
\bibitem{bib:11}A.Comtet and G.Gibbons, Nucl.Phys. B299 (1988) 719.
\bibitem{bib:12}E.Witten, Phys.Rev.Lett 38 (1977) 121.
\bibitem{bib:13}W.Nahm, Nucl. Phys. B (Proc. Suppl.) 58(1997)91.
\end{thebibliography}
\end{document}